\documentclass{elsart}

\usepackage{amssymb}
\usepackage{graphicx}

\begin{document}

\begin{frontmatter}


\title{An Implementation of the Deutsch-Jozsa Algorithm on Molecular
Vibronic Coherences Through Four-Wave Mixing: a Theoretical Study}

\author{Zsolt Bihary}
\address{Department of Chemistry, University of California, Irvine,
California 92697}
\ead{bihary@chili.ps.uci.edu}

\author{David R. Glenn}
\address{Physics Department, University of Toronto, 
60 St. George Street, Toronto, Ontario M5S 3H6, Canada}
\ead{dglenn@chem.utoronto.ca}

\author{Daniel A. Lidar\corauthref{cor1}}
\corauth[cor1]{Corresponding author. Fax: +1-416-946-7705}
\ead{dlidar@chem.utoronto.ca}
\address{Chemical Physics Theory Group, University of Toronto, 
80 St. George Street, Toronto, Ontario M5S 3H6, Canada}

\author{V. Ara Apkarian}
\ead{aapkaria@uci.edu}
\address{Department of Chemistry, University of California, Irvine,
California 92697}

\maketitle

%%%%%\title{}

%\author[label1,label2]{}
%\address[label1]{}
%\address[label2]{}

%\author{}

%\address{}

\begin{abstract}
% Text of abstract
Time-Frequency Resolved Coherent Anti-Stokes Raman Scattering
(TFRCARS) was recently proposed as a means to implement quantum logic
using the molecular ro-vibrational manifold as a quantum register
[R. Zadoyan {\it et al.}, Chem. Phys. 266, (2001) 323]. We give
a concrete example of how this can be accomplished through 
an illustrative algorithm that solves the 
Deutsch-Jozsa problem. We use realistic molecular parameters to
recognize that, as the problem size expands, shaped pulses must be
tailored to maintain fidelity of the algorithm.  
\end{abstract}

\end{frontmatter}

% main text
\section{Introduction}
Logic operations for quantum computation (QC) between two
or more quantum bits (qubits) have so far been experimentally
demonstrated in liquid-state NMR \cite{Cory:00}, trapped ions
\cite{Sackett:00}, and 
cavity quantum electrodynamics \cite{Rauschenbeutel:99}. Recently,
Time-Frequency Resolved Coherent Anti-Stokes Raman Scattering
(TFRCARS) in the molecular ro-vibronic Hilbert space, was proposed as
an alternative approach \cite{Zadoyan:01}. Through measurements on
room temperature iodine vapor, it was demonstrated that the elements
sufficient for executing universal quantum logic, namely, the single
qubit rotations and the two-qubit controlled-not gate, are naturally
contained in TFRCARS. Here, we give an illustrative example of how
such elements can be combined into a useful algorithm. We show how a
TFRCARS {\em interference} experiment can use the molecular
vibrational level structure to solve the Deutsch-Jozsa (DJ) problem
\cite{Deutsch:92}. The DJ problem has become a benchmark for
demonstrations of algorithms on prototypes of quantum
computers; having so far been used in NMR \cite
{Chuang:98a,Jones:98c,Linden:99b,Dorai:01,Marx:00,Kim:00}, linear 
optics \cite{Takeuchi:00}, ro-vibrational molecular wavepackets in a
pump-probe experiment \cite{Vala:01}, and excitons in semiconductor
quantum dots \cite{Chen:01}. 

TFRCARS is a four-wave mixing experiment in which a sequence of three
non-collinear, short laser pulses (labeled P, S and P') are used to
resonantly prepare and manipulate molecular ro-vibronic
coherences. The P-pulse acts on a statistical state to prepare the
first-order polarization
$P^{(1)} = \sum c(\xi,\xi') |\xi \rangle \langle \xi'|$ where $\xi$
represents the rotational, vibrational and electronic quantum indices
$(j,v,\chi)$. The S and P' pulses then allow manipulation of the
coherence 
through stimulated transitions to sequentially prepare second- and 
third-order coherences, consisting of complex superpositions
determined by 
the radiation field. Transform-limited 
interrogation of the evolving third-order coherence can then be
accomplished through the projective measurement of gated detection of
the spectrally dispersed third-order polarization, $P^{(3)}$
\cite{Zadoyan:01,Mukamel:book,Zadoyan:00}. Energy and momentum
conservation conditions, enforced through spectral and spatial
filtering of $P^{(3)}$, allow its detection without any interference
from 
the incoherent background. The process can be mapped out as logic
operations on the molecular register of eigenstates, by identifying
the P pulse as initialize, S and P' pulses as process, and the gate
pulse 
as readout. The approach has several attractive features with regard
to QC. Since transitions involve electronic resonances, they can be
carried out on fs time scales. A very large ratio of coherence to
process time $(>10^{5})$ is afforded, even when we allow for
decoherence to be determined by the non-fundamental limit of
inhomogeneous translational distribution (doppler broadening) at room
temperature. Unlike NMR, which aims to manipulate effective pure
states, in TFRCARS the signal is only from the polarized sub-ensemble
\cite{Zadoyan:01}. Very large superpositions, $10^2-10^4$
ro-vibrational eigenstates, can
be manipulated with precision.  Finally, since the signal consists of
a coherent radiation beam, information can be transferred effectively
and can be used to cascade operations. Small scale CARS-QC
experiments, using several qubits, are currently being set up at
Irvine. 
The limited number of qubits and process steps in the presently
conceived algorithms, suggest applications in quantum communication
\cite{DiVincenzo:00}, and simple quantum information processing
tasks, such as quantum privacy amplification and cryptography
\cite{Deutsch:96}. Even the simplest of propositions in quantum
control of molecules \cite{Rice:Book} must contend with
complications inherent in molecular state structures, such as
vibrational level structures imposed by anharmonicity of electronic
potential energy surfaces. As such, principles of optimal control 
\cite{Peirce:88} will
invariably be required in devising pulse shapes tailored for various
implementations
\cite{Tesch:01,Kosloff:02}. We highlight these considerations,
by using realistic molecular parameters, intuitively obvious pulse
shapes, and a clear computational task of solving the DJ
problem. Finally, we note that in the 
short time after suggesting the mapping of 
4-wave mixing onto QC \cite{Zadoyan:01}, there have been
demonstrations of the general approach by other groups as well 
\cite{Naumov:01,Lozovoy:02}. 

\section{Deutsch-Jozsa algorithm}
The DJ algorithm is one of the first examples of a
quantum algorithm exhibiting a speedup compared to classical computers. The
algorithm solves the following problem: ``There is a finite set $
X=\{x_{k}\}_{k=1}^{N}$ where $N$ is even, and an unspecified function $
f:X\rightarrow \{0,1\}$ that is promised to be either constant [$f(x_{k})=0$
or $1$ $\forall k$] or balanced [$f(x_{k})=0$ on exactly half the inputs].
Decide with certainty whether $f$ is constant or balanced in the least
number of evaluations of $f$.'' The DJ algorithm has been extensively
analyzed \cite{Cleve:98}. 
We now describe an alternative algorithm for solving the
DJ\ problem, one that is optimized for an interference experiment. Let
$\mathcal{H}$ 
be an $N$-dimensional Hilbert-space, where $N=|X|$. Define a basis in
$\mathcal{H}$ in one-to-one correspondence with the elements
$x_{k}\epsilon X$: $\{|x_{1}\rangle ,|x_{2}\rangle ,...,|x_{N}\rangle \}$.

\begin{enumerate}
\item
Prepare the equal
coherent superposition $|\Psi _{1}\rangle =\sum_{k=1}^{N}|x_{k}\rangle
$.
\item
Define an operator $F_{f}$ by its action on the basis elements as: $
F_{f}|x_{k}\rangle =(-1)^{f(x_{k})}|x_{1}\rangle $. Then: $|\Psi
_{2}\rangle =F_{f}|\Psi _{1}\rangle =\sum_{k=1}^{N}{(-1)^{f(x_{k})})|x_{1}
\rangle }$.
\item
Make a projective measurement on $|x_{1}\rangle $.
\end{enumerate}

The output
(the signal) becomes: 
\begin{equation}
S_{N}(f)=\langle x_{1}|\Psi _{2}\rangle =\sum_{k=1}^{N}{(-1)^{f(x_{k})}}
\label{eq:Sn}
\end{equation}
Since $f(x_{k})$ is either $0$ or $1$, the terms in the sum are $+1$
or $-1$, respectively. It is clear that for $f$ balanced, the sum
contains an equal 
number of positive and negative terms. The signal in this case is zero. For
constant $f$, all the terms have the same sign, so the absolute value of the
signal becomes maximal. The signal therefore distinguishes the two types of
functions and hence this modified algorithm performs the same task as the
standard algorithm \cite{Cleve:98}. Furthermore, it too does so using just
one $f$-evaluation.\footnote{Our algorithm is similar to one described
in \cite{Collins:98}, except that it does not use a 
qubit tensor-product structure.
}

\section{Scheme of proposed experiment}
We now propose a concrete experiment
on iodine (I$_{2}$) vapor to perform our alternative DJ algorithm. We
illustrate the proposal through numerical simulations using accurately
known molecular parameters. Vibrational levels represent the basis
states $|x_{k}\rangle $ of $|\Psi _{1}\rangle $. Fig.~1 shows the
iodine potentials for the ground ($X$) and excited ($B$) 
electronic states. We choose pulse widths and processing times $\le$~1~ps,
on which time scale the evolution of molecular rotations can be
ignored, and therefore, we need only to be concerned with the
vibrational levels on each potential. The CARS process is illustrated
in Fig.~1 through the relevant time-circuit diagram (Feynman diagram
in inset). 

Initially the
molecule is in its ground electronic 
state~$|X,0\rangle \langle 0,X| $, where the 
first (second) index denotes electronic (vibrational) quantum
numbers. 

The P (pump) pulse induces the transition
$\sum_i |B,w_i\rangle \langle 0,X| \leftarrow |X,0\rangle \langle
0,X|$. 
A short enough P-pulse is
chosen to prepare the desired number of $w$-states with comparable
amplitude. This step is the preparation of the equal coherent
superposition state $|\Psi _{1}\rangle $.

After a delay of an integer multiple of the vibrational period on the
B state, the S
(Stokes) pulse arrives to induce the transition 
$ \sum_j |X,v_j\rangle \langle 0,X| 
\leftarrow \sum_i |B,w_i\rangle \langle 0,X| $.
The delay is kept short (once or twice the vibrational period of the B
state), so the S pulse acts on the same superposition as that prepared
by the P-pulse. Otherwise, the S-pulse must be shaped to compensate
for dispersion of the vibrational packet due to the anharmonicity of
the B-state. 
The S
pulse is designed to encode the function $f$. This is done by
phase-shifting its spectral components (colors), corresponding to
the $(B,w) \rightarrow (X,4) $ transitions (or
$x_{k}$), using a phase-mask. We need only consider $0$ shifts and $\pi
$ shifts here, i.e., the mask element multiplies the spectral
component $x_{k}$ of the pulse by $(-1)^{f(x_{k})}$. 
A mask with a specific sequence of phase factors $
((-1)^{f(x_{1})},...,(-1)^{f(x_{N})})$ is therefore in one-to-one
correspondence with a function $f$. 
Note, the S pulse provides a many-to-one mapping of the B-state
vibrations on the X-state vibrations, and generates a broad
coherent superposition $ \sum_j |X,v_j\rangle \langle 0,X|$. 
However, only the 
$|X,4\rangle \langle 0,X| $ coherence carries the processed phase
information. This step corresponds to the application of 
$F_{f}$, i.e., the computation of $|\Psi
_{2}\rangle $. 

The amplitude 
transferred to the $|X,4\rangle \langle 0,X|$ coherence is measured by
applying the P'(probe) pulse, and then collecting the time-integrated
spectrally dispersed anti-Stokes radiation over a pre-selected
transition. For the P' pulse which induces the 
$\sum_k |B,w_k\rangle \langle 0,X| \leftarrow 
 \sum_j |X,v_j\rangle \langle 0,X|$ 
transition, it is sufficient to select pulse characteristics that
provide a one-to-one map of vibrational levels. Thus, for P' centered
on the $(X,4) \rightarrow (B,22) $ transition, as long as
the spectral width of the pulse is narrower than the vibrational
spacing near $v$~=~22 of 85~cm$^{-1}$
(therefore a P' pulsewidth longer than 0.7 ps),
we are guaranteed the unique
projection $(X,4) \rightarrow (B,22) $. The desired
signal is therefore obtained by dispersing the anti-Stokes radiation
through a monochromator to detect the spectral amplitude on the 
$(B,22) \rightarrow (X,0) $ transition. It is useful to
note here that by limiting the detection window at the rotational band
origin, where rotational recursions occur on a timescale $t>$~10~ps,
it 
is possible to eliminate all potential interferences from rotational
evolution. 

Let us summarize the scheme: 

1)
Preparation: The P pulse prepares the equal coherent superposition. 

2)
Computation: The phase-masked S pulse performs the $F_{f}$ operation. 

3)
Measurement: The $|X,4\rangle \langle 0,X|$ coherence is measured 
as the time integrated spectrally-dispersed CARS signal on the 
$(B,22) \rightarrow (X,0) $ transition, after application
of the vibrationally selective probe pulse. This signal is
proportional to the absolute value of the coherence $|X,4\rangle
\langle 0,X|$. 
The output characterizes $f$.

\section{Pulse design}
We require an $|X,0\rangle \rightarrow |B,w\rangle
\rightarrow |X,v\rangle $ transition scheme with the following
properties: a) For the general DJ algorithm $w$ should assume $N$ values.
Here we consider for definiteness the case of $N=4,8$ (equivalent to two and
three qubits). b) The Franck-Condon (FC) overlap between $X$ and $B$
should be as large as 
possible and roughly constant for all ($0,w$) and ($v,w$) pairs,
in order to maximize the efficiency of coherent transfer by the pulses. c)
By choosing a large frequency shift between the P and S pulses it is
guaranteed that the single time-circuit diagram of Fig.~1 describes the CARS
process. The following values satisfy the above criteria: $v=4$; $
w=20-23$ for $N=4$, and $w=18-25$ for $N=8$. The product of FC\ factors
characterizes the overall strength of the $|X,0\rangle \rightarrow
|B,w\rangle \rightarrow |X,4\rangle $ transitions. The product is largest
and roughly constant around $w=22$, which motivates our choice of the $w$
values. In Fig.~2, we show the chosen pulse-shapes for P and S in the
frequency domain for $N=4$. P is broad, covering $w=20-23$. S is
red-shifted, and phase-masked. Fig.~2 shows the effect of the $(1,-1,1,-1)$
mask, corresponding to a balanced function~$f$. Both the P pulse, and
the S pulse before masking have durations of 
$50$ fs. 
The P' pulse 
which must provide vibrational selectivity, is the same color as S but
its duration is taken to be 1~ps.

\section{Simulations}
We use Morse functions for the $X$ and $B$ electronic
potentials with parameters $D_e^X$=12550 cm$^{-1}$, 
$r_e^X$=2.666 \AA, $\beta^X$=1.858\AA$^{-1}$, 
$D_e^B$=4500 cm$^{-1}$, 
$r_e^B$=3.016 \AA, $\beta^B$=1.850\AA$^{-1}$, the energy difference
between the minima of the potentials is 15647 cm$^{-1}$
(see Fig.~1). 
We calculate the vibrational eigenfunctions on both
surfaces using the sinc-DVR method, and obtain the FC factors
\cite{Colbert:92}. The time-evolution is explicitly integrated using
the 
energy-representation. Time-ordering of the P and S pulses was not
enforced: contributions from (P,S) and (S,P) sequences were added
coherently. Only the Liouville pathway depicted in Fig.~1 was taken
into 
account. No decoherence mechanism was included (decoherence, which
occurs with
$\tau > 10^{-9}s$, can be neglected on the execution
time-scale of 1-2 ps). After the application of the two pulses, the
prepared $ 
|X,4\rangle \langle X,0|$ vibrational coherence
(which is an explicit functional of the phase mask and the delay
between the P and S pulses)
preserves its magnitude $A(f,\tau ) = ||X,4\rangle \langle X,0|| $
and oscillates with the $(X,0)-(X,4)$ beat-frequency. 
By explicit calculation of the 
time-integrated CARS spectrum, using the perturbative evaluation of 
$P^{(3)}(t)$ as described in detail in Ref.~\cite{Zadoyan:01}, 
we verify that the experimentally observable
$(B,22) \rightarrow (X,0) $ spectral component 
is proportional to $A(f,\tau ) $. In what follows we simply consider
$A(f,\tau ) $ as the signal.

\section{Results}
The signal $A(f,\tau )$ as a function of delay-time $\tau $
between the P and S pulses was computed for all $2^{N}$ different
phase-masks (i.e., for all 
possible different encoded $f$ functions, including those that are neither
constant nor balanced). Fig.~3 shows the $N=4$ signals 
as functions of the delay time between the P and S pulses
for a constant and a
balanced function. For the constant function (left) the signal is large at
integer multiples of the ($B$) period, while the signal is close to zero for
the balanced function (right). This is therefore a clear experimental
signature that distinguishes constant from balanced functions. Notice,
however, that the signal from the balanced function is not exactly zero. In
Fig.~4a, we show the signal $A(f,1)$ for all $16$ different encoded $f$
functions {\it vs.} $S_{4}(f)$ [Eq.~(\ref{eq:Sn})] (only $8$ points are
actually shown, since the signal is symmetric in $\pm S_{4}(f)$). For
balanced functions $S_{4}(f)=0$, so they should yield zero signal. For the
two constant functions $|S_{4}(f)|=4$, so they should give maximal signal.
For the other functions $|S_{4}(f)|=2$, so they should give half the signal
amplitude. In reality, as expected from Fig.~3, there is a spread along the
signal axis, indicating that not all balanced functions yield exactly
zero signal. Fig.~4b shows the results for $N=8$ case. Here the spread of
signals corresponding to the same class of functions is significantly
larger, thus reducing the fidelity of distinguishing constant from
balanced functions. It is useful to introduce a formal fidelity measure in
order to quantify the performance of the algorithm. Perhaps the simplest
measure is the distinguishability $D$, which we define as the ratio of
signal strengths

\begin{equation}
D=1-\frac{\max A(f_{{\rm bal}})}{A(f_{{\rm con}})},
\end{equation}
where the maximum is taken over all balanced signals. In the ideal case all $
A(f_{{\rm bal}})=0$ so $D=1$. A second measure is the Pearson's correlation
coefficient $r$ for a straight line interpolation between the signal $
A(f,\tau )$ and $|S_{N}(f)|$, which also accounts for the spread in the
other (neither constant nor balanced)\ functions. This measure is important
in case one decides to use our algorithm to also distinguish these other
functions. In Table~1 we present these fidelities for different $N$ values
and for different delay times. The reasons leading to a degradation in
fidelity are the following: I) Non-uniform FC factors, causing different $
|x_{k}\rangle $ to be transferred with slightly different amplitude by the $P
$ and S pulses. II) Spectral breadth of the laser-pulses, with the same
consequence as I). These effects lead to non-uniformity in the prepared and
projected coherent superpositions, thus misrepresenting $|\Psi _{1}\rangle $
and $|\Psi _{2}\rangle $ somewhat. III) Anharmonicity of the $(B)$
potential, leading to wavepacket dispersion on both the $X$ and $B$
potentials. In order to explore these effects, we performed
simulations for the $N=8$ 
case, where all FC factors were artificially set equal (a pulse achieving
this effect can in principle be shaped by amplitude masking), and the
P pulse was spectrally 
broadened, by reducing its duration to $10$ fs. We refer to this as the
``tailored'' case. The results are presented in Fig.~4c, and the improvement
in distinguishability is remarkable. This convincingly demonstrates the role
of the non-uniform superposition in fidelity degradation. The solution, a
spectrally broader P pulse together with tailoring the {\em amplitudes} of
the P and S pulses to compensate for the non-uniformity of the FC
factors, greatly improves the fidelity of the algorithm. To test the role of
anharmonicity we also performed simulations at delays $\tau =0,2\tau_B$.
Obviously, at zero delay anharmonicity cannot play a role. On the other hand
anharmonicity should be more significant at $\tau =2\tau_B$ than at
$\tau =\tau_B$. 
These expectations are confirmed in Table 1: a significant improvement in
fidelity is seen when comparing the bare $N=8$ results to the tailored
results, but only for $\tau =0,1\tau_B$. In contrast, for $\tau
=2\tau_B$ the fidelity 
remains low. The degradation for large $N$
values at large delay should be attributed to wavepacket dispersion due to
anharmonicity. The simple solution is to apply the S pulse at $\tau
=0$ or $\tau_B$. The choice $\tau =\tau_B$ is actually preferable
since interference from the 
time-reversed (S,P) sequence at $\tau =0$, as well as higher order
processes (6-wave mixing etc.) not accounted for in our simulations,
lead to smaller signal 
strength. Optimization of the {\em phases} of the S and P pulses
(chirping) is another option that can be used for compensating for
anharmonicity effects \cite{Bardeen:97a}.

\section{Discussion and Conclusions}
The CARS-QC proposal \cite{Zadoyan:01}
is a promising new implementation of quantum logic, with potential near-term
implications for quantum communication. Our main purpose in this work was to
test the feasibility of this proposal by studying in detail the
implementation of a benchmark quantum algorithm. We chose the Deutsch-Jozsa
(DJ) problem for its conceptual and technical simplicity, and devised a
modified algorithm that only requires quantum interference, not
entanglement. This allowed us to directly test aspects related to the
preparation of input vibrational superposition states and their dynamical
evolution. To maintain fidelity, it is necessary to devise pulses that
compensate for non-uniform Franck-Condon factors and counteract
anharmonicity 
effects. These can be achieved through amplitude masking and frequency
chirping using spatial light modulators \cite{Weiner:95}, to
optimally tailor the fields for a given choice of delay
between pulses. Here, we have indicated that intuitively
obvious pulses already allow a clear demonstration of how
computational 
algorithms may be implemented in TFRCARS. The strategy in the
laboratory demonstration of this, and similar algorithms involving
rotation-vibration-electronic qubits, will be to rely on genetic
algorithms to design pulses optimized for particular tasks
\cite{Judson:92,Brixner:00,Wilson:97}.    

\section*{Acknowledgements}
D.A.L. gratefully acknowledges support from
NSERC. The QC effort at Irvine is supported through the US AFOSR,
under grant F49620-01-1-0449.

%\bibliographystyle{elsart-num}
%\bibliography{/home/dlidar/articles/bib}

\clearpage

\begin{table}[h]
\centering 
\begin{tabular}{|l|c|c|c|}
\hline
{$N$} & {$\tau=0$} & {$\tau=1$} & {$\tau=2$} \\ \hline
{4} & {99/86} & {99/90} & {99/85} \\ \hline
{6} & {95/79} & {95/78} & {91/69} \\ \hline
{8} & {84/60} & {84/62} & {73/41} \\ \hline
{8t} & {99/89} & {97/73} & {81/43} \\ \hline
\end{tabular}
\caption{Percentage Correlation/Distinguishability for DJ problem instances
with different $N$ and time delays $\protect\tau$. Last line (8t):
``tailored'' implementation.}
\label{CD:tbl}
\end{table}

\clearpage

\begin{figure}
\centering
\includegraphics[height=10cm,angle=270]{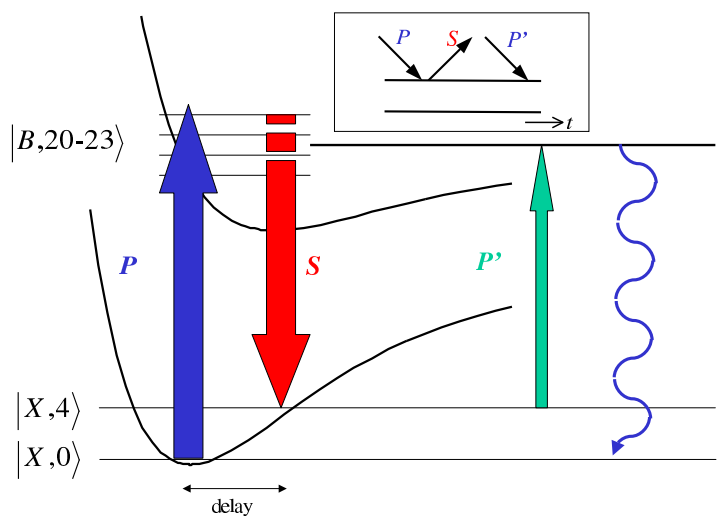}
\caption{$X$ (ground) and $B$ (excited) electronic potential energy
surfaces of iodine, modelled as Morse potentials. Also shown:
time-cicuit diagram of pump (P), Stokes (S), and probe ($P^{\prime
}$) pulses, and relevant vibrational states. Inset: Corresponding Feynman diagram}
\label{fig1}
\end{figure}

\begin{figure}
\centering
\includegraphics[height=10cm]{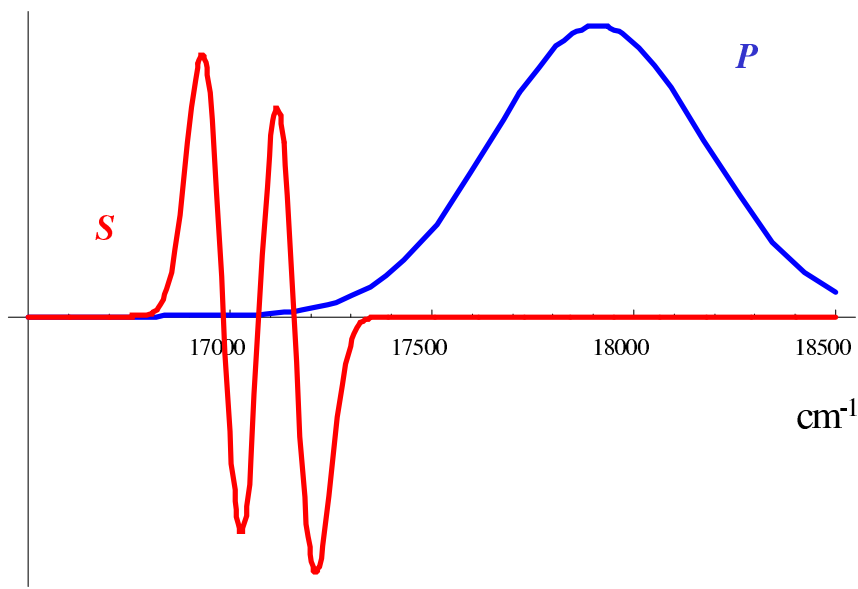}
\caption{Spectrum of P and S pulses. S pulse for (1,-1,1-1) mask is
shown.}
\label{fig2}
\end{figure}

\begin{figure}
\centering
\includegraphics[height=10cm]{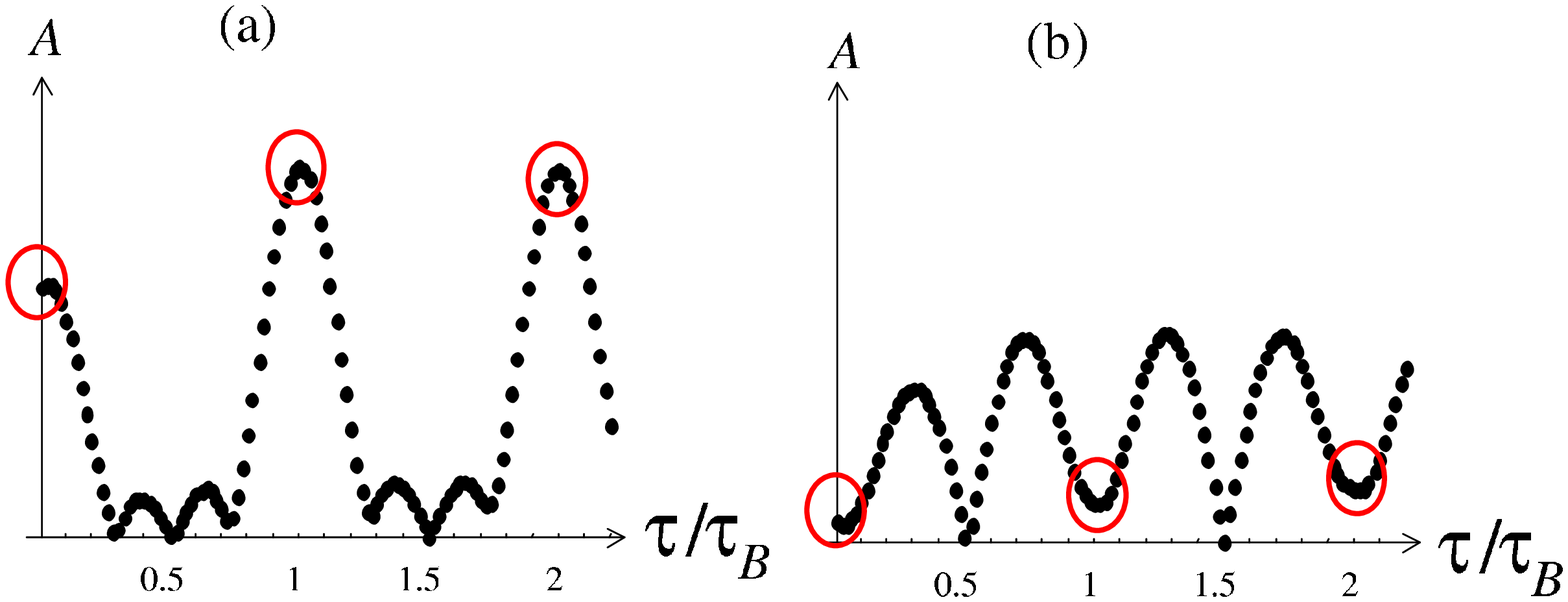}
\caption{Signal $A$ as function of delay time $\protect\tau$ for different
masks. Time in units of ($B$) vibrational period. a) Constant function. b)
Balanced function.}
\label{fig3}
\end{figure}

\begin{figure}
\centering
\includegraphics[height=10cm]{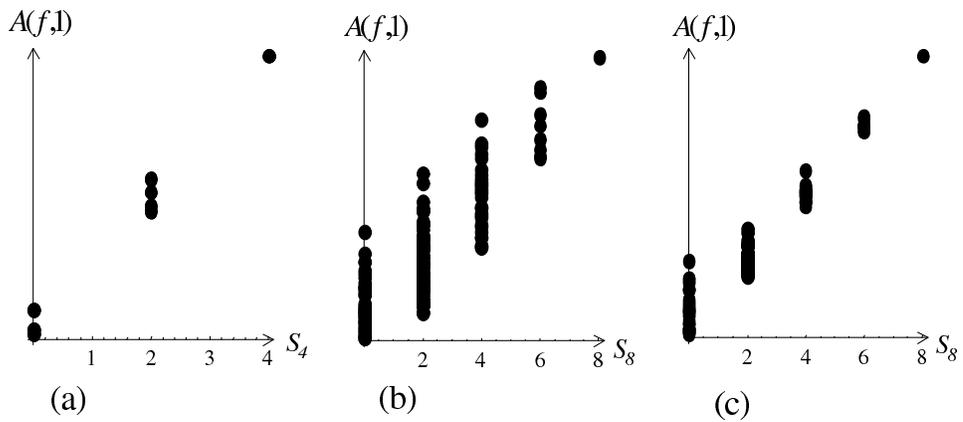}
\caption{Signal $A$ as function of $S_N(f)$. a) $N=4$, b) $N=8$, c)
``tailored'' $N=8$.}
\label{fig4}
\end{figure}

\end{document}